\documentclass[aip,reprint]{revtex4-1}
\usepackage{graphicx}% Include figure files
\usepackage{dcolumn}% Align table columns on decimal point
\usepackage{bm}% bold math
\usepackage{amsmath}
\usepackage{float}
\draft % marks overfull lines with a black rule on the right

\begin{document}
%\preprint{}

\title{Non-linear macro evolution of a dc driven micro atmospheric glow discharge} %Title of paper



\author{S. F. Xu }
\noaffiliation
\affiliation{The State Key Laboratory on Fiber Optic Local Area, Communication Networks and Advanced Optical Communication Systems, Key Laboratory for Laser Plasmas and Department of Physics and Astronomy, Shanghai Jiao Tong University, Shanghai 200240, China}
%\thanks{}
\author{X. X. Zhong}
\email{xxzhong@sjtu.edu.cn}
\noaffiliation
\affiliation{The State Key Laboratory on Fiber Optic Local Area, Communication Networks and Advanced Optical Communication Systems, Key Laboratory for Laser Plasmas and Department of Physics and Astronomy, Shanghai Jiao Tong University, Shanghai 200240, China}

\date{\today}

\begin{abstract}
We studied the macro evolution of the micro atmospheric glow discharge generated between a micro argon jet into ambient air and static water. The micro discharge behaves similarly to a complex ecosystem.  Non-linear behaviors are found for the micro discharge when the water acts as a cathode, different from the discharge when water behaves as an anode. Groups of snapshots of the micro discharge formed at different discharge currents are captured by an intensified charge-coupled device with controlled exposure time, and each group consisted of 256 images taken in succession. Edge detection methods are used to identify the water surface and then the total brightness is defined by adding up the signal counts over the area of the micro discharge. Motions of the water surface at different discharge currents show that the water surface lowers increasingly rapidly when the water acts as a cathode. In contrast, the water surface lowers at a constant speed when the water behaves as an anode. The light curves are similar to logistic growth curves, suggesting that a self-inhibition process occurs in the micro discharge. Meanwhile, the total brightness increases linearly during the same time when the water acts as an anode. Discharge-water interactions cause the micro discharge to evolve. The charged particle bomb process is probably responsible for the different behavior of the micro discharges when the water acts as cathode and anode.
\end{abstract}

\pacs{47.61.Ne, 52.80.Hc, 52.70.-m}% microscale flows, glow discharge, plasma diagnostic

\maketitle %\maketitle must follow title, authors, abstract and \pacs

\section{Introduction}
Micro discharges spatially confined to dimensions of 1 mm or less have been used in biomedicine, plasma sources , particle kinetics, nanofabrication, thin film processing, etching and deposition because of their non-equilibrium character.\cite {Sankaran2003,Kikuchi2004,Iza2008,Mariotti2012,Ostrikov2013} In particular, the micro atmospheric glow discharges generated between gas nozzle microjets into ambient air and static liquids have recently attracted considerable attention.\cite{Richmonds2008,Chiang2010,Ehlbeck2011,Patel2013}

This work was partly motivated by a previous study.\cite{Xu2015} Discharge-water interactions mean that plasma causes the water level to lower and the micro discharge to subsequently lengthen. This makes it impossible to measure a two-dimensional temperature map of the micro discharge as done in the reference. The discharge-liquid interactions cause the micro discharge to gradually change over time. However, many numerical simulations and experiments have focused on analyzing the temporal evolution of micro plasmas over a very short time scale.\cite{Wilson2008,Yatom2012,Blajan2012,Shlapakovski2015} The evolution of micro discharge on a macro time scale has not been well studied, which currently limits the application of micro discharge.

In this paper, we study the macro evolution of a micro discharge generated between an argon jet into ambient air and water at atmospheric pressure. Edge detection methods are used to identify the water surface and then total brightness is defined by adding up the signal counts over the area of the micro discharge. Light curves are extracted to describe the macro evolution of the micro discharge. Non-linear behaviors are found in the micro discharge when water acts as a cathode, different from the discharge of a water anode.
\section{Experiment}
%################
\begin{figure}
\includegraphics{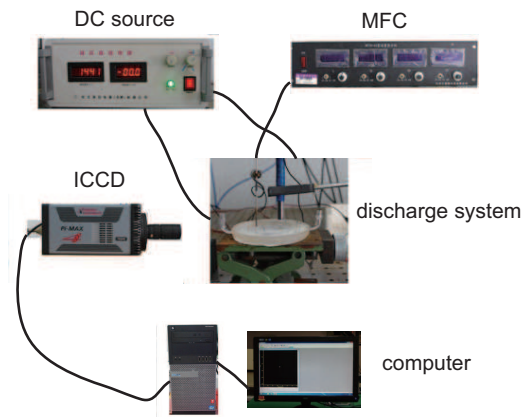}
\caption{Schematic diagram of the experimental system.}
\label{device}
\center
\end{figure}
%%%%%%%%%%%%%%%%%
The experimental system consisted of a direct current (DC) source, mass flow controller (MFC), discharge system, intensified charge-coupled device (ICCD)  and computer, as shown in Fig.~\ref{device}. The DC source provided a high voltage to drive the micro discharge. The MFC controlled the flux of argon gas through the stainless-steel capillary into the ambient air. The capillary (internal diameter $175$ $\mu m$)  also acted as an electrode. 
%##########################
\begin{figure*}
\center
\includegraphics{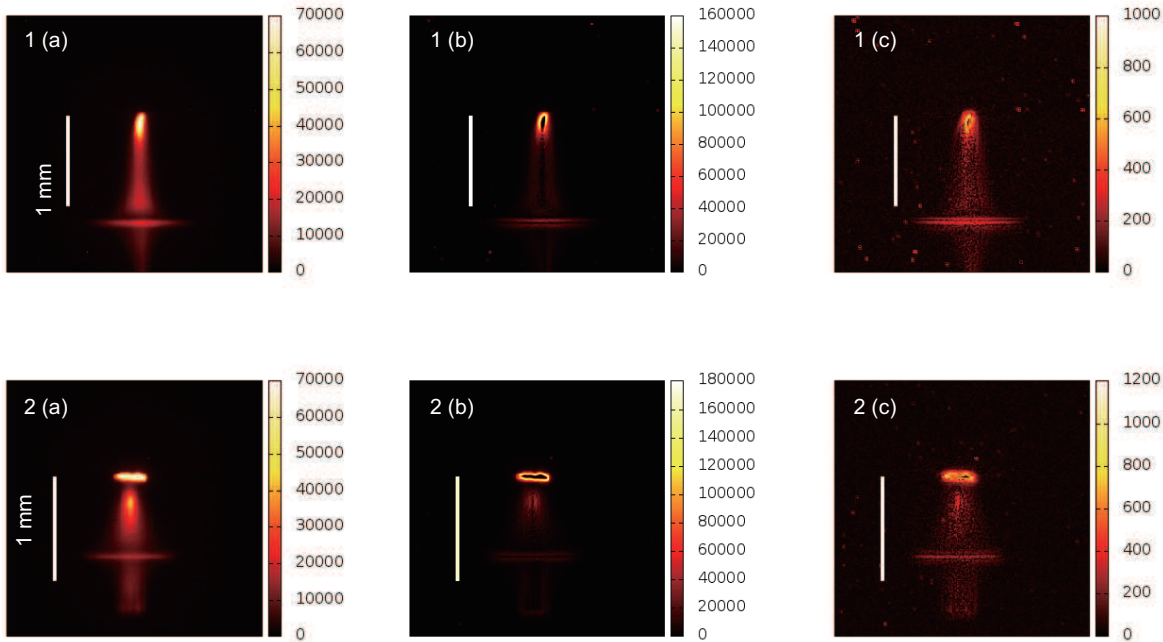}
\caption{Edge detection images . The water acts as a cathode in the first row, as an anode in the second row. In each row, the images from left to right are the source image, gradient image of the source image and divergence image of the gradient image. The discharge current was $8$ $mA$ and the exposure time was $10$ $ms$.}
\label{boundary}
\center
\end{figure*}
%##########################
The diameter of the quartz dish in the discharge system used as a water container was about $60$ mm. The radiation of the micro discharge was collected by the ICCD and then analyzed by the computer. The volume of water was fixed to 25 $mL$ and its resistance was about 1430 $\Omega$. The initial gap between the exit and water surface was adjusted to about 1 mm. The water behaved as either a cathode or anode. In each case, the adjustable control parameter was the discharge current. Because of the heavy mass of argon atoms, when we increased the gas flux to $20$ $sccm$ or higher, the strong argon jet flow entered the water and the water surface bent.  Therefore, the argon gas flux was fixed at $10$ $sccm$.

Groups of snapshots of the micro discharge at different discharge currents were taken by the ICCD with controlled exposure time and time step. Each group of snapshots consisted of 256 images taken in succession. When the water acted as a cathode, a snapshot was taken every 10 $s$ at a certain discharge current; for example, 8 $mA$. The exposure time and time step were $10$ $ms$ and $10$ $s$, respectively. We scanned discharge currents from 8 to 12 $mA$ at intervals of 1 $mA$. When the water used as an anode was changed, the above operations were repeated.

The source image $f(x,y)$ has been used to denote the snapshot of the micro discharge. The ICCD array consisted of $1024\times 1024$ pixels, so the source image $f(x,y)$ is a $1024\times 1024$ two-dimensional matrix. The count of a pixel in the source image is proportional to the radiation intensity of the micro discharge.
\section{Results and discussion }
%################
Discharge-water interactions cause the water surface to lower as time progresses. We used the edge detection method to accurately detect the level of the water surface in the micro discharge system, as illustrated in Fig.~\ref{boundary}. The principles of this method are as follows. The source image $f(x,y)$ will show a local maximum on the line of the water surface. The gradient $|\nabla f(x,y)|$ of the source image equals zero. The continuous differential operators $\frac{\partial f}{\partial x}$ and $\frac{\partial f}{\partial y}$ were replaced by difference operators. Instead of centered difference, the Sobel operation  was used. In fact, the computation process involved the convolution between the source image $f(x,y)$ and Sobel template operator.

However, the gradient of source image is not good enough to find the water surface. Because a dark region existed around the water surface, the gradient image contained three local maxima. The divergence of the gradient image also results in a local maximum at the line of the water surface. Therefore, the Gauss-Laplacian operator was used to compute triangle f(x,y) as follows
\begin{equation}
\triangle f(x,y)=\frac{\partial^2 f(x,y)}{\partial x^2}+\frac{\partial^2 f(x,y)}{\partial y^2}.
\end{equation}

\begin{figure}[H]
\center
\includegraphics[width=8 cm]{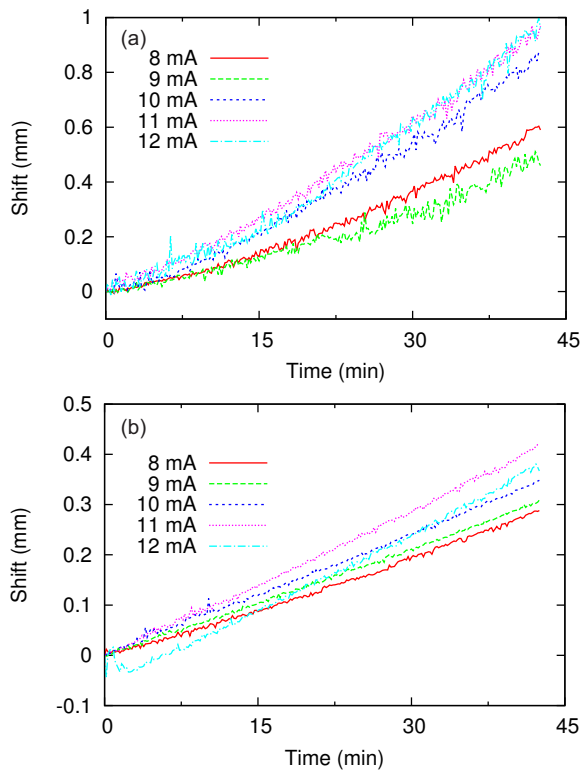}
\caption{Motion of a water surface exposed to an argon jet  at different discharge currents over 43 minutes. Water acting as (a) a cathode, and (b) an anode.}
\label{surface}
\center
\end{figure}

Fig.~\ref{surface} shows the motion of the water surface at different discharge currents. The slope of the curve represents the speed of the water surface. Obviously, the water surface lowers at a constant rate when the water acts as an anode, proving a uniform motion. However, the motion equation is similar to an exponential function when the water acts as a cathode, proving there is accelerated motion compared with when the water behaves as an anode. The motion can be described phenomenologically as:
\begin{equation}
Shift(t)=Ae^{Bt}+C
\end{equation}
where $A$, $B$ and $C$ are just mathematical parameters.

The discharge current and speed of the water surface have a strong positive correlation; a larger discharge current results in a larger speed. At the same discharge current, the speed is higher when the water acts as a cathode than when it behaves as an anode.
%%%%%%%%%%%%%%%%%
%%%%%%%%%%%%%%%%%%%
A light curve was defined as a graph of the total brightness of the micro discharge as a function of time. The total brightness of the micro discharge increased as time progressed, which was defined by adding up the effective signal counts over the area of the source image $f(x,y)$. The effective signal count was determined above the water surface when it was larger than 12 standard deviations than the average of background.

\begin{figure}
\center
\includegraphics[width=8 cm]{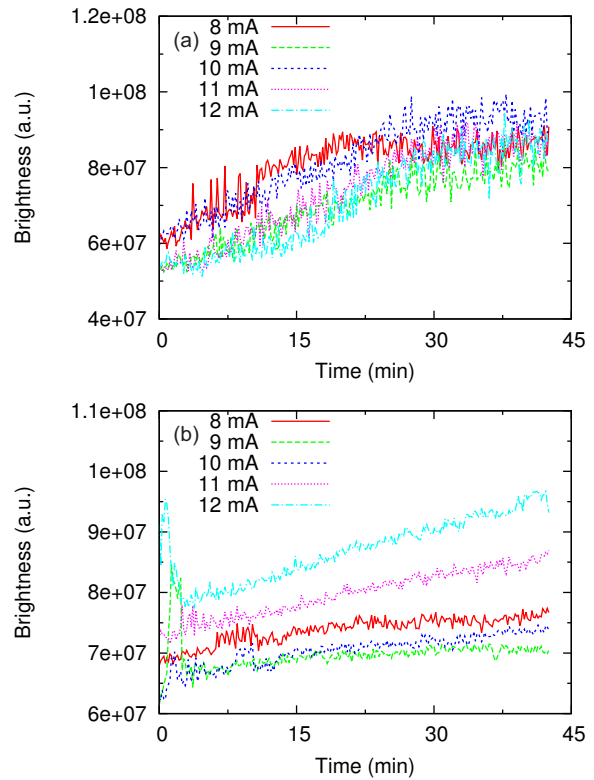}
\caption{Light curves of micro discharges at different discharge currents. for water acting as (a) a cathode, and (b) an anode.}
\label{curve}
\center
\end{figure}

Fig.~\ref{curve} displays light curves for the systems at different discharge currents. The total brightness increased slowly initially, and then increased faster. After a turning point, the growth rate begins to decrease gradually and finally becomes saturated. The shape of the light curves was similar to a logistic growth curve, suggesting there was a self-inhibition process when the water acted as a cathode. When the water behaved as an anode, there was no turning point. The total brightness increased linearly. Initially, flare occurred around the exit and greatly increased the total brightness in a short time when the water acted as an anode, as shown in Fig.~\ref{curve} (b). This flare disturbs the detection of the water surface and should be managed carefully.
%%%%%%%%%%%%%%%%%%%%%%%%%
%%%%%%%%%%%%%%%%%%%%%%%%%%

The behavior of the system is markedly different depending on whether the water behaves as a cathode or anode. That is to say, the processes associated with discharge-water interactions are different when water acts as a cathode or anode.

Liquid water dissipates in a variety of ways. First, water is consumed by electrolysis. The shift of the water surface caused by electrolysis is estimated as $Shift=\frac{MI}{2\pi r^2\rho N_A e}t$ from the Faraday law of electrolysis, where $M$ is the molar mass of water, $I$ is the discharge current, $r$ is the radius of the quartz dish, $\rho$ is the mass density of water, $N_A$ is the Avogadro constant, and $e$ is elementary charge. The speed that the water is lowered by electrolysis  is about $3.96\times 10^{-7}\ mm/s$ at the maximum discharge current of the micro discharge of $I=12\ mA$. Electrolysis has the same effect on the micro discharge at the same discharge current when the water acts as either cathode or anode.

Meanwhile, the temperature of the water increased over time, which enhanced the rate of evaporation of the liquid water. If Joule heat is all used to increase water temperature, the estimated temperature change is $\triangle T=\frac{I^2R}{mc}t$, where $t$ is time, $I$ is the discharge current, $R$ is the water resistance, $m$ is the mass of water, and $c$ is the specific heat capacity of water. After 2562.56 s, the temperature of the water is increased by 5 ¡ãC at the maximum discharge current of $I=12\ mA$. Therefore, Joule heating is not effective in this system.

A more effective process is energy transport  from the discharge to the water. It is known that the temperature of neutral gas in micro discharge is much higher than the boiling point of water at standard atmospheric pressure.\cite{Lu2013,Xu2013} Therefore, micro discharge is able to heat the water at the discharge-water interface. High-temperature discharge flows across the water surface, and greatly increases the water surface temperature around the discharge-water interface.

There is also an indirect vaporization process in the system. The energetic charged particles of the micro discharge accelerated by the applied electric field bomb  the water surface. These particles sputter small water droplets or water clusters from the liquid water into the high-temperature micro discharge. These small water droplets evaporate instantly. When the water acts as a cathode, positive ions bomb the water surface. Conversely, when the water acts as an anode, negative ions bomb the water surface. Because the discharge was in air and there was discharge-water interaction, the positive species included $Ar^{+}$, $N_2^{+}$ and so on, and the negative species included electrons, $NO_2^{-}$, $NO_3^{-}$ and so on.\cite{Nikiforov2011,Li2012} The density of positive species was roughly equivalent to the density of negative species. Even though the mass of an argon ion is smaller than that of heavy complex negative ions ($NO_2^{-}$, $NO_3^{-}$), the density of argon ion is much larger than that of heavy negative ions. Therefore, positive ions are more powerful than negative ions when bombing the water surface.

\begin{figure*}
\center
\includegraphics{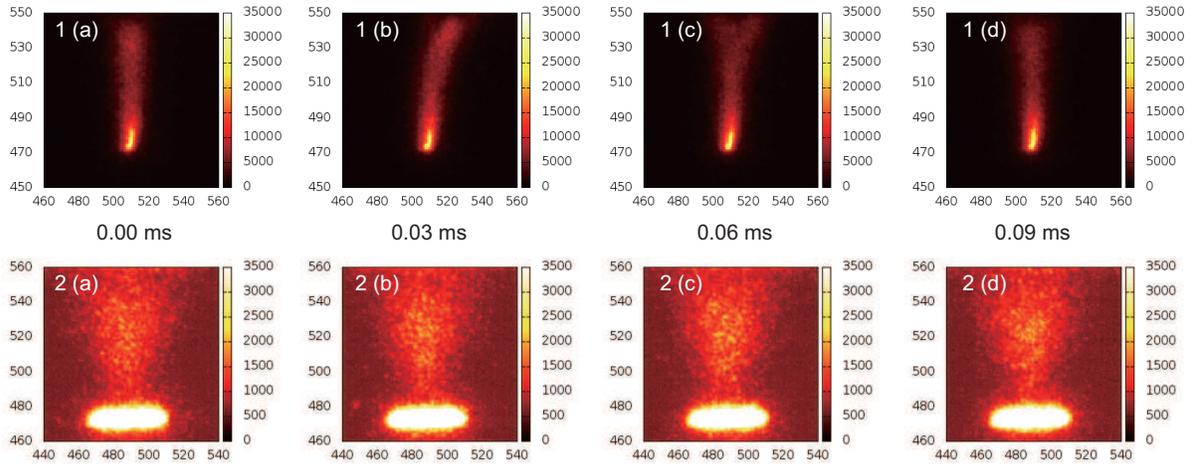}
\caption{Snapshots of micro discharge at different times from 0.00 to 0.09 $ms$. In the first row, the water acts as a cathode. In the second row, the water acts as an anode. The exposure time was $0.01$ $ms$. The horizontal and vertical coordinates represent the pixel position.}
\label{vibrate}
\center
\end{figure*}

Shortening the exposure time to $0.01$ $ms$ and the time step of the ICCD to $27$ $ns$, we obtained a group of snapshots revealing information about the rapid changes of the micro discharge over a short time scale. Fig.~\ref{vibrate} shows there are major differences between the plasma-liquid interactions when water acts as a cathode and anode. Obvious vibration and instability exist in the discharge luminance, suggesting that plasma-liquid interactions are stronger when the water acts as a cathode than an anode.
%####################

\section{Conclusion}
Micro atmospheric glow discharge was generated between an argon microjet into ambient air and static water. We studied the macro evolution of the micro discharge using an ICCD with high dynamic range. Non-linear evolution and differences in behavior between the cases where water behaves as a cathode and anode were found.

Edge detection methods were originally introduced to detect the water surface from source images of the micro discharges. The water surface lowered more rapidly as time passed when the water acted as a cathode. Conversely, the water surface lowered at a constant speed when the water acted as an anode.

The micro discharge and its total brightness behaved similarly to ecosystem and biotic populations.  The light curves of the micro discharge resembled logistic growth curves when the water acted as a cathode, suggesting there was a self-inhibition process in the micro discharge. The interactions between the micro discharge and water, especially charged particle bomb process, play important roles in the evolution of the micro discharge.

The optical spectrum emitted by the micro plasma also changed correspondingly over time. How does the change of the microplasma emission spectrum influence the total brightness and its instability? A more detailed study is required to answer this question.
\begin{acknowledgments}
XXZ and SFX acknowledge support from the NSFC (Grant No. 11275127, 90923005), STCSM (Grant No. 09ZR1414600) and MOST  of China. We thank Kostya Ostrikov for editing the manuscript.
\end{acknowledgments}

\providecommand{\noopsort}[1]{}\providecommand{\singleletter}[1]{#1}%

\end{document}